\documentclass[11pt]{article} 
\usepackage{setspace}
\usepackage{pdflscape}
\usepackage[utf8]{inputenc} 
\usepackage{amsmath}
\usepackage{amssymb}
\usepackage{endnotes}
\usepackage{amsmath,amsthm,amssymb,amsfonts,mathrsfs}
\usepackage[colorlinks,
linkcolor=blue,
anchorcolor=blue,
citecolor=blue,
]{hyperref}

\usepackage{geometry} 
\usepackage{graphicx} 

\usepackage{booktabs} 
\usepackage{array} 
\usepackage{paralist} 
\usepackage{verbatim} 
\usepackage{subfig} 
\usepackage{amsmath} 
\usepackage{fancyhdr} 
 \usepackage[numbers,sort&compress]{natbib} 
\usepackage{makecell}
\usepackage{multirow}
\usepackage{sectsty}
\allsectionsfont{\sffamily\mdseries\upshape} 
\usepackage{indentfirst}
\usepackage[nottoc,notlof,notlot]{tocbibind} 
\usepackage[titles,subfigure]{tocloft} 
\usepackage{authblk}
\usepackage{abstract}

\title{Convective meta-thermal dispersion for self-adaptive cooling enhancement}
\author{Xinchen Zhou\textsuperscript{1,2}, Ruzhu Wang\textsuperscript{2}, Xiaoping Ouyang\textsuperscript{3,*}, Jiping Huang\textsuperscript{1,*}}
\affil{\textsuperscript{1}Department of Physics, State Key Laboratory of Surface Physics, Key Laboratory of Micro and Nano Photonic Structures (MOE), Fudan University, Shanghai 200438, China;}
\affil{\textsuperscript{2}School of Mechanical Engineering, Institute of Refrigeration and Cryogenics, Engineering Research Center of Solar Power and Refrigeration (MOE), Shanghai Jiao Tong University, Shanghai 200240, China;}
\affil{\textsuperscript{3}School of Materials Science and Engineering, Xiangtan University, Xiangtan 411105,
China.}

\affil{*To whom correspondence should be addressed; E-mail: oyxp2003@aliyun.com, jphuang@fudan.edu.cn.}
\begin{document}
\maketitle
\setstretch{1.1}
\begin{abstract}

Improving the heat transfer coefficient is crucial across various energy utilization processes for maintaining device safety and stability with high energy efficiency. However, in scenarios with limited heat capacity flow rates, increasing the thermal conductivity of encapsulated internal heat source (IHS) packaging can paradoxically impede heat transfer. Herein, we introduced a convective-meta thermal dispersion (CMTD) strategy applicable throughout the energy domain. By integrating low thermal conductivity materials into high thermal conductivity package structures, we disrupted tangential heat flow while preserving efficient radial heat transport. Through this approach, a notable reduction in tangential temperature within the fluid channel was achieved, effectively lowering the IHS temperature. Remarkably, this cooling mechanism does not need additional energy input, thermal property enhancements, or expanded heat transfer areas, which are often prerequisites in existing technologies. Moreover, spontaneous enhancement phenomena emerged under constrained heat transfer conditions, termed self-adaptive cooling enhancement. Our investigations revealed, under steady-state conditions, a maximum 24.5\% decrease in IHS average temperature, while transient conditions exhibited a maximum 32.3\% increase in heat transfer between the IHS and cooling fluid, validating the efficacy of the CMTD strategy. These findings offer a promising pathway for efficient thermal management in various thermal energy utilization fields with high power density such as nuclear fission and fusion and contributed to a deeper understanding of fundamental fluid-solid heat transfer mechanisms across the energy science.
\end{abstract}

\section{Introduction}
Optimal temperature control plays a pivotal role in the operation of critical systems, ranging from battery systems and electronic devices to nuclear power plants \cite{heenan2023mapping,van2020co,lei2022prediction}. The imperative of ensuring the safety and stability of these intricate systems is heightened against the backdrop of contemporary imperatives such as energy conservation and emissions reduction \cite{van2020co}. To meet these challenges head-on, it is imperative to harness advanced technologies that facilitate efficient thermal management.

In the overarching heat transfer process between the core units that generate heat in real-time, termed the internal heat source (IHS), and the working fluids, the utilization of fluid-solid heat transfer stands as a prevalent method \cite{sarkar2014analytical}. Within this framework, the fluid traverses through the solid package structure encapsulating the IHS, engaging in heat exchange. The imperative to minimize energy input necessitates the maintenance of an efficient heat transfer mechanism. Adhering to Newton's cooling law, which delineates the heat transfer rate between a solid surface and the flowing fluid, the improvement of convective heat transfer coefficients, enlargement of heat transfer area, or enhancement of thermal properties emerge as widely acknowledged strategies. This multifaceted approach, known as heat transfer enhancement, commands significant attention and ongoing research efforts \cite{van2020co, DIACONU2023106830, liu2022heat, MOUSA2021110566}.

While various research endeavors achieve ideal heat transfer enhancement based on active or passive methods such as vibration \cite{UNNO2020119588, ABADI201997}, external magnetic fields \cite{BEZAATPOUR2020114462, BhattacharyyaSuvanjan2022, JAFARI2020106495}, pulsating flow \cite{KUMAVAT2022107790, YeQianhao2021}, enlarging roughness \cite{LIU2022117850, PrakashChandra2019}, adding high conductivity additives \cite{ZHANG2022280, WEI2021100948, SaadoonZahraaH.2022}, secondary flow \cite{ParamanandamKarthikeyan2022, SADIQUE2022123063}, etc., there is a lack of study focused on the temperature distribution of the working fluid and its complex interaction with the IHS temperature, which might help develop advanced thermal management technologies \cite{HE2021119223}. Specifically, the comprehensive influence of the thermal properties of the package structure on the thermal management performance of the IHS remains unclear. To address this issue, we examine a heat transfer model, illustrated in Fig. 1a. At the core of this model lies an IHS encased in an annular package structure, surrounded by a cooling fluid. The fluid moves from left to right, exchanging heat with the package structure. When the fluid velocity reaches a sufficient magnitude, its temperature can be considered constant. A higher thermal conductivity ($\kappa_\mathrm{s}$) of the package structure results in a lower temperature of the IHS (Fig. 1b). However, the velocity of the cooling fluid is often restricted in scenarios like thermal energy utilization for nuclear reactors. The temperature of the working fluid varies along the flow channel, and a portion of the heat flows tangentially from the IHS to the working fluid, increasing the tangential temperature distribution of the cooling fluids. This tangential flow does not effectively contribute to the cooling performance of the IHS. Then, an extraordinary phenomenon is observed: reducing $\kappa_\mathrm{s}$ leads to a lower IHS temperature, contrary to intuition (Fig. 1c). This suggests that simply increasing $\kappa_\mathrm{s}$ as a conventional strategy cannot enhance the thermal management performance of the IHS. While existing material science can achieve ultra-high thermal conductivity up to $2000\sim5000\ \mathrm{W}\ \mathrm{m}^{-1}\ \mathrm{K}^{-1}$ \cite{ge2022graphene, WANG2023101182, WANG2023124598}, the temperature variation characteristics of the working fluid in the flow channel restrict its effectiveness. Thus, conducting fundamental research on the fluid-solid heat transfer process is crucial, not only for understanding the temperature control mechanism but also for technological advancement.
\begin{figure}[htpb!]
\centering
\includegraphics[width=\textwidth]  {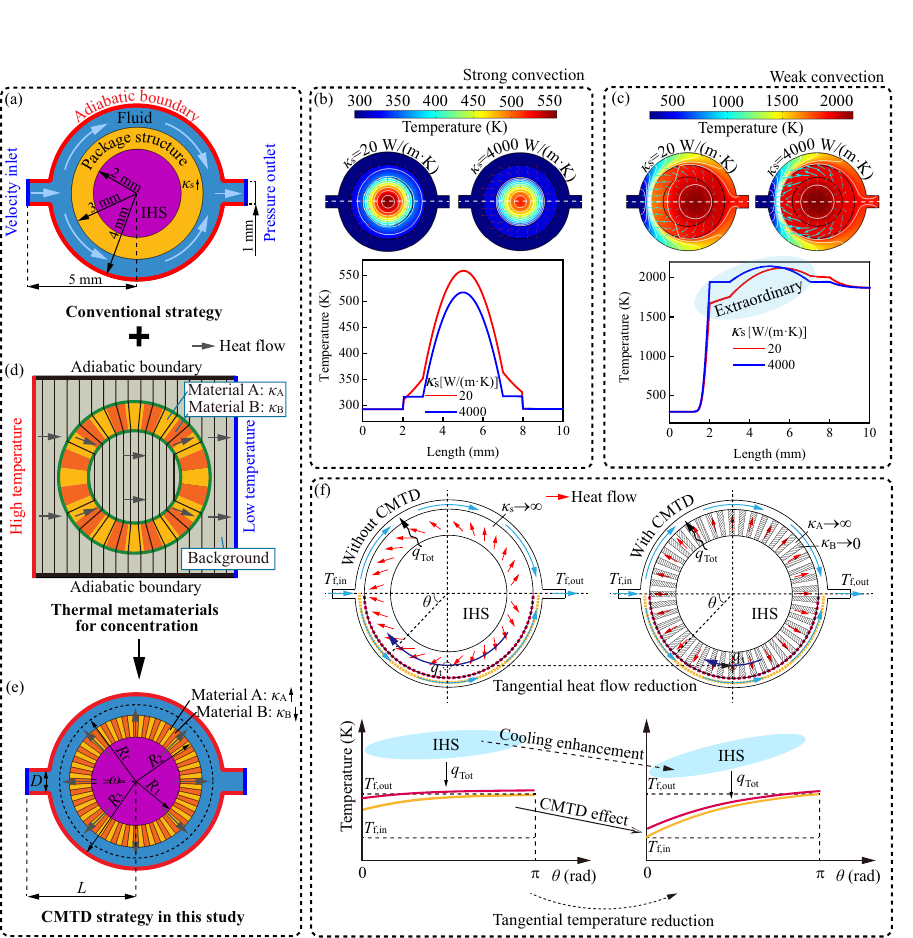}
\caption{\label{Fig. 1} 
\footnotesize{CMTD strategy for self-adaptive cooling enhancement. (a) Heat transfer model for an internal heat source (IHS) packaged with single isotropic material under fluid cooling conditions (model abstraction of conventional strategy). The left and right depict velocity inlet and pressure outlet boundaries, respectively. The upper and lower boundaries are adiabatic. (b,c) Thermal management performances under (b) strong and (c) weak convection. The upper subgraphs show temperature distributions with low and high thermal conductivity package structures ($\kappa_\mathrm{s}=20\ \mathrm{W\ m^{-1}\ K^{-1}}$ and $4000\ \mathrm{W\ m^{-1}\ K^{-1}}$), while the lower subgraphs illustrate temperature distribution along the central line of the heat transfer model as marked by the white dashed line. (d) Schematic of thermal concentration enabled by thermal concentrators, a kind of thermal metamaterials, using two materials with high and low thermal conductivity ($\kappa_\mathrm{A}$ and $\kappa_\mathrm{B}$). (e) Heat transfer model with two isotropic materials with high and low thermal conductivities to achieve equivalent high radial and low tangential thermal conductivities (CMTD strategy in this study). (f) CMTD effect on regulating the heat flow in the package structure for realizing cooling enhancement of the IHS. $T_\mathrm{f,in}$ and $T_\mathrm{f,out}$ represent the inlet and outlet temperatures of the cooling fluid, respectively; $q_\mathrm{Tot}$ denotes the heat transfer rate from the IHS to the cooling fluid.
}}
 \end{figure}

Addressing this concern, recent advancements in thermal metamaterials and related diffusive manipulation technologies offer an efficient method for regulating heat flow in both engineering applications and theoretical tools \cite{huang2020theoretical, xu2023transformation, FanCZ, LiY, ShenXY, YangS, XuL, XUL2, LiYNRM, LiYNM, JinPPNAS, zhang2023diffusion, xu2023giant, zhou2023adaptive, jin2023deep, yang2023controlling, D3EE03214K}. This is achieved by comprehensively considering the thermal and structural parameters of the heat transfer system. For instance, a specific type of thermal metamaterial, known as a thermal concentrator, has been developed to concentrate heat flow into the region inside the shell (functional area, Fig. 1d), thereby improving thermal harvesting efficiency \cite{han2015manipulating, li2022energy, han2013theoretical}. Existing research in this field has primarily focused on fundamental aspects under both heat conduction and convection, beginning with analytical theory and translating into real-world applications through effective medium approximation \cite{xu2023black, JinPPNAS, zhuang2022breaking}.

Drawing on the structural concept of thermal concentrators in thermal metamaterials, we apply it to the dispersion of heat from the IHS to the cooling fluid through convection heat transfer, termed convective meta-thermal dispersion (CMTD, Fig. 1e). This allows us to regulate the tangential heat flow (Fig. 1f), resulting in a decrease in tangential fluid temperature and the outer surface temperature of the package structure, consequently reducing the IHS temperature. This approach involves incorporating low thermal conductance into the original high thermal conductivity package structure to optimize convective cooling effects, which significantly diverges from conventional methods focusing on thermal property enhancement and heat transfer area enlargement.

In this study, we unveil the impact of CMTD on the cooling enhancement performance of the IHS across various thermal and structural parameters under both steady and transient conditions, while delving into its underlying heat transfer mechanism. Furthermore, we observe and analyze several self-adaptive cooling enhancement phenomena occurring under constrained heat transfer conditions. Our discoveries and conclusions may pave the way toward achieving efficient thermal management and foster a deeper understanding of fluid-solid heat transfer mechanisms.

\section{Results and discussions}

\subsection{CMTD for self-adaptive cooling enhancement: Mechanism and effects}

We illustrate the mechanism of CMTD using the IHS-based heat transfer model under a specific case (Fig. 2a, Methods). The designed CMTD aims to lower the IHS temperature by regulating the heat flow within the package structure connecting the IHS and the cooling fluid, resulting in a dispersion phenomenon observed from the inset of the case without CMTD to that with CMTD. Specifically, the original heat flow along the tangential direction turns to the radial direction under the action of CMTD.

 \begin{figure}[h!]
\centering
\includegraphics[width=\textwidth]  {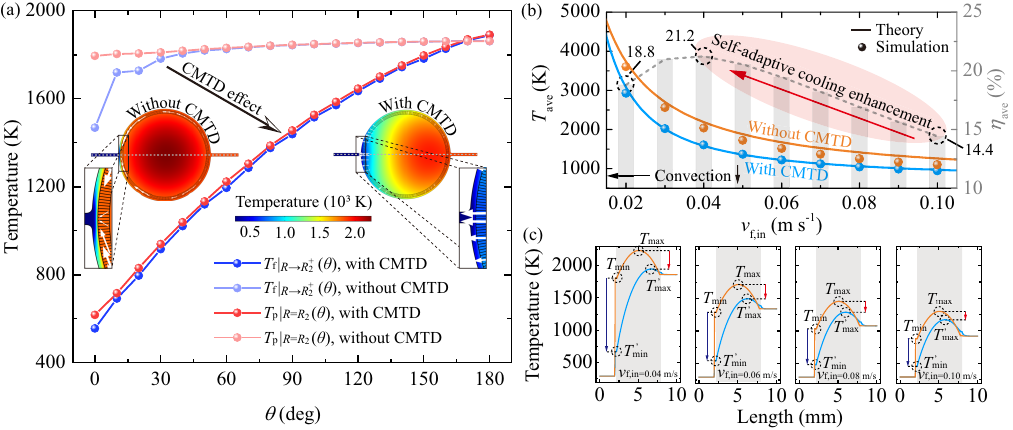}
\caption{\label{Fig. 2} 
\footnotesize{Mechanism of CMTD on self-adaptive cooling enhancement. (a) Temperature distributions of the fluid $\big(T_\mathrm{f}|_{R\rightarrow R_2^{+}}\big)$ and the outer surface of the package structure $\big(T_\mathrm{p}|_{R=R_2}\big)$ along the tangential direction with and without CMTD. The insets are the temperature distributions of the heat transfer model with and without CMTD. The white arrows represent heat flow. (b) The relationships between the average temperature of the IHS ($T_\mathrm{ave}$) and the inlet velocity of cooling fluid ($v_\mathrm{f,in}$). (c) Temperature distributions of the central lines (the white lines marked in the inset of (a)) in the heat transfer model under the conditions of $v_\mathrm{f,in}=\{0.04, 0.06, 0.08, 0.10\}\  \mathrm{m\  s^{-1}}$.
}}
 \end{figure} 

We established two datasets to examine the effects of CMTD. One dataset comprises the temperature distributions of the fluid positioned close to the outer surface of the package structure, marked as $T_\mathrm{f}|_{R\rightarrow R_2^{+}}(\theta)$. The other dataset represents the temperature distributions of the outer surface of the package structure, signified as $T_\mathrm{p}|_{R= R_2}(\theta)$. In the case without CMTD, high thermal conductivity materials are used to prepare the package structure, facilitating easy heat transportation along the tangential direction, leading to higher $T_\mathrm{f}|_{R\rightarrow R_2^{+}}(\theta)$ and $T_\mathrm{p}|_{R= R_2}(\theta)$, and even causing $T_\mathrm{p}|_{R= R_2}(0)\rightarrow T_\mathrm{p}|_{R= R_2}(\pi)$ (Fig. 2a).

As depicted in Fig. 1e, CMTD includes low thermal conductivity materials to break the tangential heat flow and maintain high radial heat transportation by the high thermal conductivity materials. Under this operation, $T_\mathrm{f}|_{R\rightarrow R_2^{+}}(\theta)$ and $T_\mathrm{p}|_{R= R_2}(\theta)$ significantly decrease, leading to the cooling enhancement within the temperature distributions of the IHS (Fig. 2a).

We then observed a phenomenon in $T_\mathrm{f}|_{R=R_2^{+}} (\theta)$ enabled by CMTD. In cases without CMTD, $T_\mathrm{f}|_{R=R_2^{+}} (0)$ is close to $T_\mathrm{f}|_{R=R_2^{+}} (\pi)$, while in cases with CMTD, $T_\mathrm{f}|_{R=R_2^{+}} (0)$ is close to $T_\mathrm{f,in}$. We conducted a theoretical analysis to solve the temperature fields of the IHS (Supplementary Note 1, Supplementary Figs. 1 and 2). As a representative result, the average temperature of the IHS ($T_\mathrm{ave}$) is given by
\begin {equation}
\left\{\begin{aligned}
T_\mathrm{ave}&=\left\{\left(\frac{1}{6\kappa_\mathrm{h}}+\frac{1}{2 h R_2}+\frac{1}{2\kappa_\mathrm{r}}\mathrm{ln}\frac{R_2}{R_1}+\frac{\pi d_\mathrm{z}}{c_\mathrm{f}\dot m}\right) R_1^2\right\}\dot\phi_\mathrm{h}+T_\mathrm{f,in}\\
T_\mathrm{ave}'&=\left\{\frac{2aR_1}{9\kappa_\mathrm{h}}+\frac{R_1^2}{2hR_2}+\frac{R_1^2}{2\kappa_\mathrm{r}}\mathrm{ln}\frac{R_2}{R_1}+\frac{\left(a+R_1\right)^2}{36\kappa_\mathrm{h}}\left(5+\frac{K\left[-\frac{4aR_1}{\left(R_1-a\right)^2}\right]}{E\left[-\frac{4aR_1}{\left(R_1-a\right)^2}\right]}\right)\right\}\dot\phi_\mathrm{h}+T_\mathrm{f,in}
\end{aligned}
\right.,
\end {equation}
where $T_\mathrm{ave}$ and $T_\mathrm{ave}^{'}$ represent the average temperatures of the IHS without and with CMTD, respectively; $\kappa_\mathrm{h}$ denotes the thermal conductivity of the IHS; $h$ is the convective heat transfer coefficient between the package structure and the cooling fluid; $\kappa_\mathrm{r}$ is the (equivalent) radial thermal conductivity of the package structure; $R_1$ and $R_2$ are the inner and outer radii of the package structure, respectively; $\dot{\phi}_\mathrm{h}$ is the power density of the IHS; $c_\mathrm{f}$ and $\dot{m}$ are the specific heat capacity and mass flow rate of the cooling fluid, respectively; $\dot{m}$ can be further expressed as $\dot{m}=\rho_\mathrm{f,in}v_\mathrm{f,in}D d_\mathrm{z}$, where $D$ is the inlet and outlet flow channel size (Fig. 1e); $d_\mathrm{z}$ denotes the thickness of the heat transfer model; $K[\cdots]$ and $E[\cdots]$ are the complete elliptic integrals of the first and second kinds, respectively; $a$ is the distance from the highest temperature point of the IHS to the geometric center of the heat transfer model, which reads $a=\pi R_1\kappa_\mathrm{h}d_\mathrm{z}/\left(c\dot{m}\right)$ (Supplementary Note 1, Supplementary Fig. 1b). Based on the convective heat transfer correlation of the heat transfer model (Supplementary Notes 2 and 3, Supplementary Tables 1 and 2), we plotted the relationships between the average temperature of the IHS and the fluid inlet velocity (Fig. 2b, Supplementary Note 4, Supplementary Figs. 3-5). The agreement of theoretical and simulation results quantitatively demonstrates that the temperature reduction of the IHS is due to the thermal dispersion within the package structure between the IHS and the cooling fluid.

Additionally, we defined cooling enhancement rates to illustrate the extent of the reduction in the IHS temperature facilitated by CMTD compared to cases without it (Methods). Corresponding to the relationship between $T_\mathrm{ave}$ and $T_\mathrm{ave}^{'}$, we depict the correlation between the average cooling enhancement rate $\eta_\mathrm{ave}$ and $v_\mathrm{f,in}$. A phenomenon is observed wherein, as $v_\mathrm{f,in}$ decreases from $0.10\ \mathrm{m\ s^{-1}}$ to $0.02\ \mathrm{m\ s^{-1}}$, $\eta_\mathrm{ave}$ initially increases from 14.4\% to 21.2\%, then decreases to 18.8\%. A more detailed temperature distribution of the central lines of the heat transfer model is presented in Fig. 2c. It reveals an extreme value for the CMTD effect on reducing IHS temperature with the variation of $v_\mathrm{f,in}$, illustrating that when the convection heat transfer intensity decreases within a certain range, the CMTD becomes significantly effective. In the range of $0.04-0.10\ \mathrm{m\ s^{-1}}$, the lower the inlet velocity, the stronger the cooling enhancement rate. Since this phenomenon occurs spontaneously, we term it a self-adaptive cooling enhancement.

\subsection{Steady cooling enhancement characteristics under different thermal and structural parameters}

Building upon the heat transfer principle depicted in Fig. 1e, we delve deeper into the factors influencing self-adaptive cooling enhancement within a segment of the heat transfer model (Fig. 3a). The effectiveness of CMTD relies on the redistribution of fluid temperature fields facilitated by the radial and tangential heat flux ($q_\mathrm{r}$ and $q_\mathrm{t}$) within the package structure. Initially, we explore the anisotropic nature resulting from the alternating distribution of materials A and B. To achieve this, we conducted 15 distinct cases, each featuring varied combinations of materials A and B for CMTD (Fig. 3b). Cases numbered 1-14 depict scenarios with CMTD in effect, while case 15 represents the absence of CMTD. We computed the corresponding equivalent radial and tangential thermal conductivities, as well as the degree of anisotropy (Methods, Supplementary Note 4). Notably, as the case number decreases, the thermal conductivity of material A ($\kappa_\mathrm{A}$) remains constant, while the thermal conductivity of material B ($\kappa_\mathrm{B}$) gradually diminishes from $4000 \ \mathrm{W\ m^{-1}\ K^{-1}}$ to $1 \ \mathrm{W\ m^{-1}\ K^{-1}}$, indicating a progressive increase in the degree of thermal conductivity anisotropy ($\varXi$). Leveraging the heat transfer process delineated in Fig. 1e, we conducted steady-state simulations, accounting for an equivalent level of anisotropy alongside varying thermal and structural parameters of the heat transfer model (Methods).

 \begin{figure}[h!]
\centering
\includegraphics[width=\textwidth]  {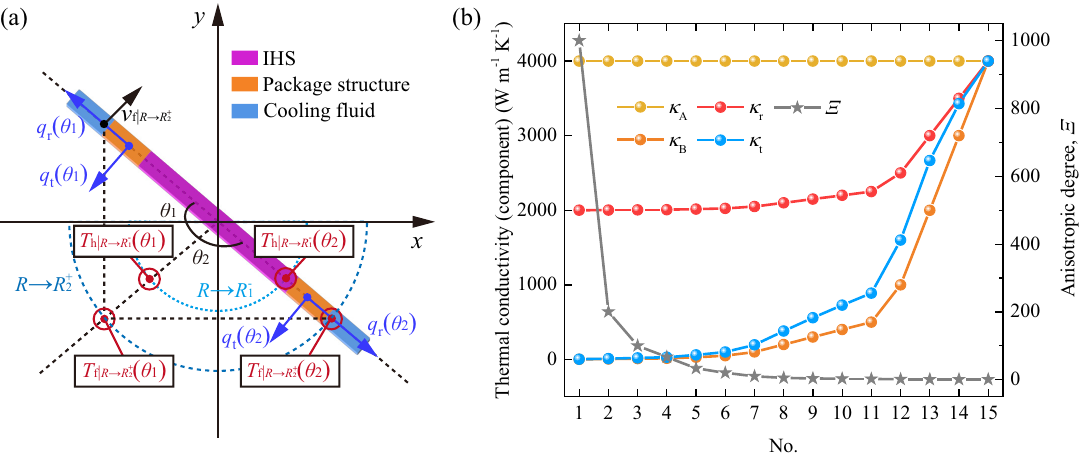}
\caption{\label{Fig. 3} 
\footnotesize{Anatomy of the factors affecting macroscopic convective heat transfer. (a) The profile of the heat transfer model. $q_\mathrm{r}\left(\theta_1\right)$ and $q_\mathrm{r}\left(\theta_2\right)$ [$q_\mathrm{t}\left(\theta_1\right)$ and $q_\mathrm{t}\left(\theta_2\right)$] denote the radial [tangential] heat flux in the package structure at positions $\theta_1$ and $\theta_2$, respectively. $T_\mathrm{f}|_{R\rightarrow R_2^{+}}\left(\theta_1\right)$ and $T_\mathrm{f}|_{R\rightarrow R_2^{+}}\left(\theta_2\right)$ [$T_\mathrm{h}|_{R\rightarrow R_1^{-}}\left(\theta_1\right)$ and $T_\mathrm{h}|_{R\rightarrow R_1^{-}}\left(\theta_2\right)$] represent the tangential temperature of the cooling fluid [IHS] close to the package structure $R\rightarrow R_2^{+}$ [$R\rightarrow R_1^{-}$] at positions $\theta_1$ and $\theta_2$, respectively. $v_\mathrm{f}|_{R\rightarrow R_2^{+}}$ denotes the tangential velocity of the cooling fluid close to the package structure. (b) Thermal conductivity (components) setting of package structure. $\kappa_\mathrm{t}$ is the equivalent tangential thermal conductivity of the package structure. See details in Methods, Supplementary Note 4, and Supplementary Table 3.
}}
 \end{figure} 
 
Moreover, we characterized the tangential velocity and temperature distributions, both with and without CMTD, of the cooling fluid and IHS near the package structure [$v_\mathrm{f}|_{R\rightarrow R_2^{+}}\left(\theta\right)$, $T_\mathrm{f}|_{R\rightarrow R_2^{+}}\left(\theta\right)$, and $T_\mathrm{h}|_{R\rightarrow R_1^{-}}\left(\theta\right)$] to quantitatively analyze the effect of CMTD on regulating their temperature fields. Cases with CMTD exhibit appropriate $v_\mathrm{f}|_{R\rightarrow R_2^{+}}\left(\theta\right)$, which significantly regulate $T_\mathrm{f}|_{R\rightarrow R_2^{+}}\left(\theta\right)$ and further reduce $T_\mathrm{h}|_{R\rightarrow R_1^{-}}\left(\theta\right)$ compared to cases without CMTD, indicating the effectiveness of thermal dispersion in achieving cooling enhancement. For instance, considering the tangent plane shown in Fig. 3a, a notable CMTD effect is evidenced by a substantial improvement in [$T_\mathrm{f}|_{R\rightarrow R_2^{+}}\left(\theta_2\right)$-$T_\mathrm{f}|_{R\rightarrow R_2^{+}}\left(\theta_1\right)$] and [$T_\mathrm{h}|_{R\rightarrow R_1^{-}}\left(\theta_2\right)$-$T_\mathrm{h}|_{R\rightarrow R_2^{-}}\left(\theta_1\right)$], highlighting the capability of CMTD in disrupting $q_\mathrm{t}$ while maintaining high $q_\mathrm{r}$.

\begin{figure}[h!]
\centering
\includegraphics[width=\textwidth]  {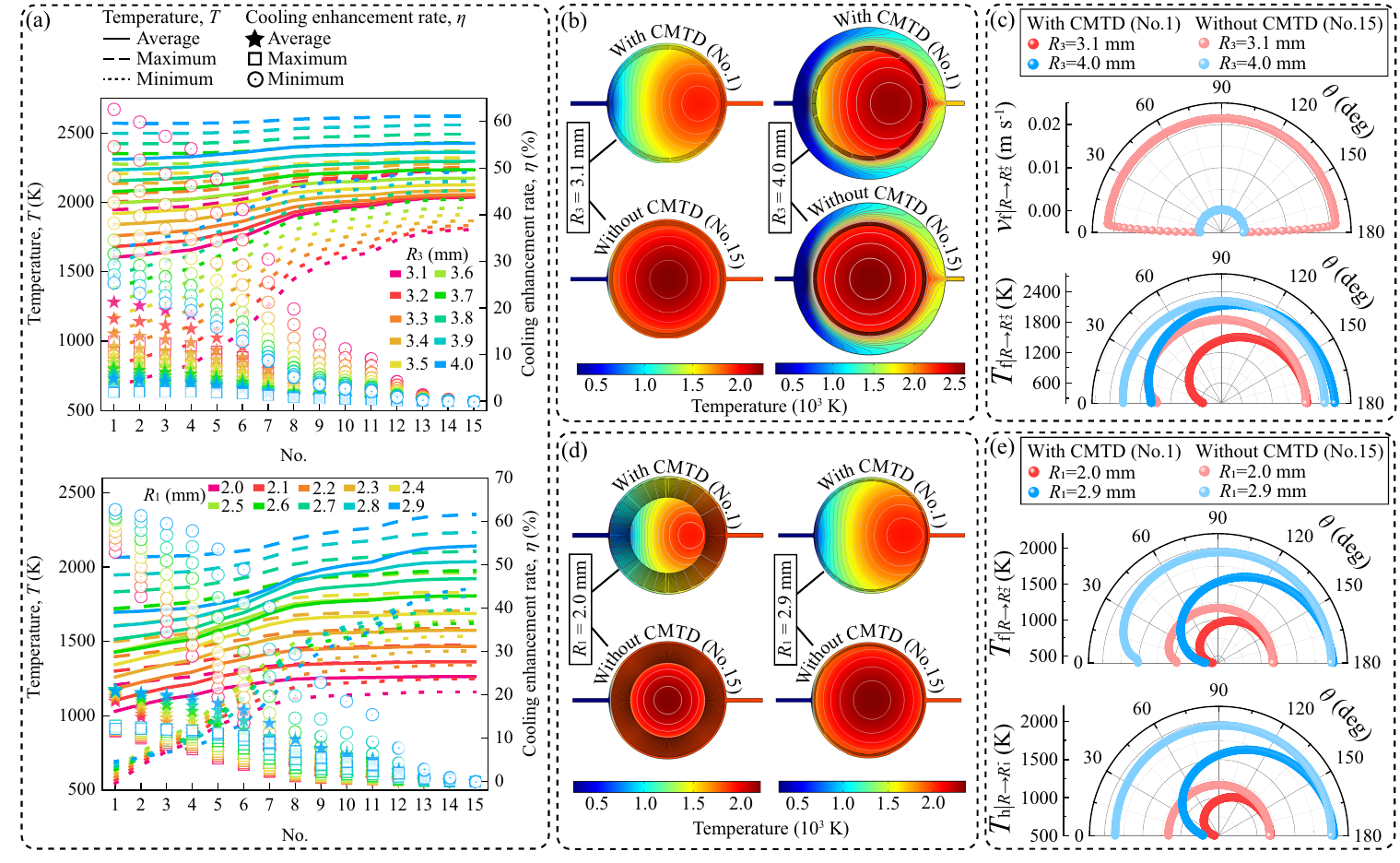}
\caption{\label{Fig. 4}
\footnotesize{Steady cooling enhancement characteristics under different structural parameters. (a) The maximum/average/minimum temperature of the IHS and the maximum/average/minimum cooling enhancement rate of the case with and without CMTD. The upper and lower subgraphs consider the variations of $R_3$ and $R_1$, respectively. (b-e) Temperature distributions of the heat transfer model and the tangential temperature/velocity characterizations. (b,d) Temperature distributions of the heat transfer model with and without CMTD under the conditions $R_3=3.1\  \mathrm{mm}$, $R_3=4.0\  \mathrm{mm}$ (b); $R_1=2.0\  \mathrm{mm}$, and $R_1=2.9\  \mathrm{mm}$ (d). (c) Tangential velocity distributions $(v_\mathrm{f}|_{R\rightarrow R_2^{+}})$ and temperature distributions $(T_\mathrm{f}|_{R\rightarrow R_2^{+}})$ of the working fluid close to the package structure for (b). (e) Tangential temperature distributions $(v_\mathrm{f}|_{R\rightarrow R_2^{+}})$ of the working fluid and the tangential IHS temperature $(T_\mathrm{h}|_{R\rightarrow R_1^{-}})$ close to the package structure for (d).}}
 \end{figure}

Figures 4-6 present the maximum, average, and minimum temperatures of the IHS under both CMTD and non-CMTD conditions, alongside the corresponding cooling enhancement rates, considering various structural, fluidic, and IHS parameters. Since self-adaptive cooling enhancement relies on the thermal dispersion capability within the package structure, a stronger equivalent anisotropic degree yields more significant cooling efficiency. An intriguing phenomenon is evident in Figs. 4a, 5a, and 6a: as the case number decreases from 15 to 1, and thus, the thermal conductivity of material B diminishes, the IHS temperature decreases instead of increasing, leading to a notable improvement in cooling performance. This is attributed to the fact that the reduced radial thermal conductivity, always higher than $2000 \ \mathrm{W\ m^{-1}\ K^{-1}}$, adequately facilitates heat transportation along the radial direction (Fig. 3b). Meanwhile, the tangential thermal conductivity undergoes significant reduction, curbing unnecessary tangential heat transfer and thereby lowering the tangential temperature of the cooling fluid. This phenomenon, observed across various thermal and structural parameters, further underscores the impact of CMTD on enhancing IHS cooling across broader contexts.

As the self-adaptive cooling enhancement is attributed to the temperature redistribution of the fluid in the flow channel enabled by CMTD, the detailed heat transfer relationships between the IHS and cooling fluid contribute strong correlations with the cooling enhancement rate. We begin to elucidate this relationship by investigating the influence of structural parameters and inserting the fluidic and IHS parameters to deepen the mechanism.

We begin by examining the CMTD effect under varying channel sizes, keeping $R_2$ constant while altering $R_3$ from $3.1\ \mathrm{mm}$ to $4.0\ \mathrm{mm}$ (Fig. 4a, upper subgraph). As $R_3$ decreases, the flow channel narrows ($R_3-R_2$ decreases), leading to a higher cooling enhancement effect (Fig. 4b). To understand this phenomenon, we analyze the tangential velocity and temperature distributions close to the package structure ($v_\mathrm{f}|_{R\rightarrow R_2^{+}}$ and $T_\mathrm{f}|_{R\rightarrow R_2^{+}}$, Fig. 4c). The variation in the flow channel size significantly influences $v_\mathrm{f}|_{R\rightarrow R_2^{+}}$ (Fig. 4c, upper subgraph). Theoretical analysis indicates strong correlations between the IHS temperature distribution and convective heat transfer intensity, impacting the redistribution effect (Eq. (1), Supplementary Note 1). The markedly enhanced redistribution effect before and after CMTD from $R_3=4.0\ \mathrm{mm}$ to $R_3=3.1\ \mathrm{mm}$ suggests an optimal tangential velocity distribution for maximizing the CMTD effect (Fig. 4c, lower subgraph), aligning with observations in Fig. 2b.

Further, we focus on the influence of inlet velocity on the CMTD effect (Fig. 5a). This analysis complements Fig. 2b. The optimal inlet velocity $v_\mathrm{f,in}=0.04\ \mathrm{m\ s^{-1}}$ corresponds to maximum $\eta_\mathrm{min}$, $\eta_\mathrm{ave}$, and $\eta_\mathrm{max}$. The CMTD effect diminishes away from $v_\mathrm{f,in}$, as shown by cases under $v_\mathrm{f,in}=0.01\ \mathrm{m\ s^{-1}}$ and $v_\mathrm{f,in}=0.10\ \mathrm{m\ s^{-1}}$ (Fig. 5b,c). We generalize this phenomenon to the magnitude of heat capacity flow rate $C_\mathrm{f}$, defined as $C_\mathrm{f}=c_\mathrm{f}\rho_\mathrm{f}v_\mathrm{f,in}Dd_\mathrm{z}=c_\mathrm{f}\dot m$. It reflects the fluid's ease in receiving the IHS's influence to change temperature. We then examine the role of fluid density $\rho_\mathrm{f}$ in this mechanism (Fig. 5a,d,e). By varying $\rho_\mathrm{f}$ from $200\ \mathrm{kg\ m^{-3}}$ to $5000\ \mathrm{kg\ m^{-3}}$, we find the maximum cooling enhancement rates occur at $\rho_\mathrm{f}=2000\ \mathrm{kg\ m^{-3}}$. Cooling enhancement rates decrease significantly away from $\rho_\mathrm{f}=2000\ \mathrm{kg\ m^{-3}}$, indicating a suitable relationship between $\rho_\mathrm{f}$ and the CMTD effect. Further tests on varying fluid heat capacity $c_\mathrm{f}$ yield similar results, underscoring the significant impact of $C_\mathrm{f}$ on the CMTD effect (Supplementary Fig. 6).

 \begin{figure}[h]
\centering
\includegraphics[width=\textwidth]  {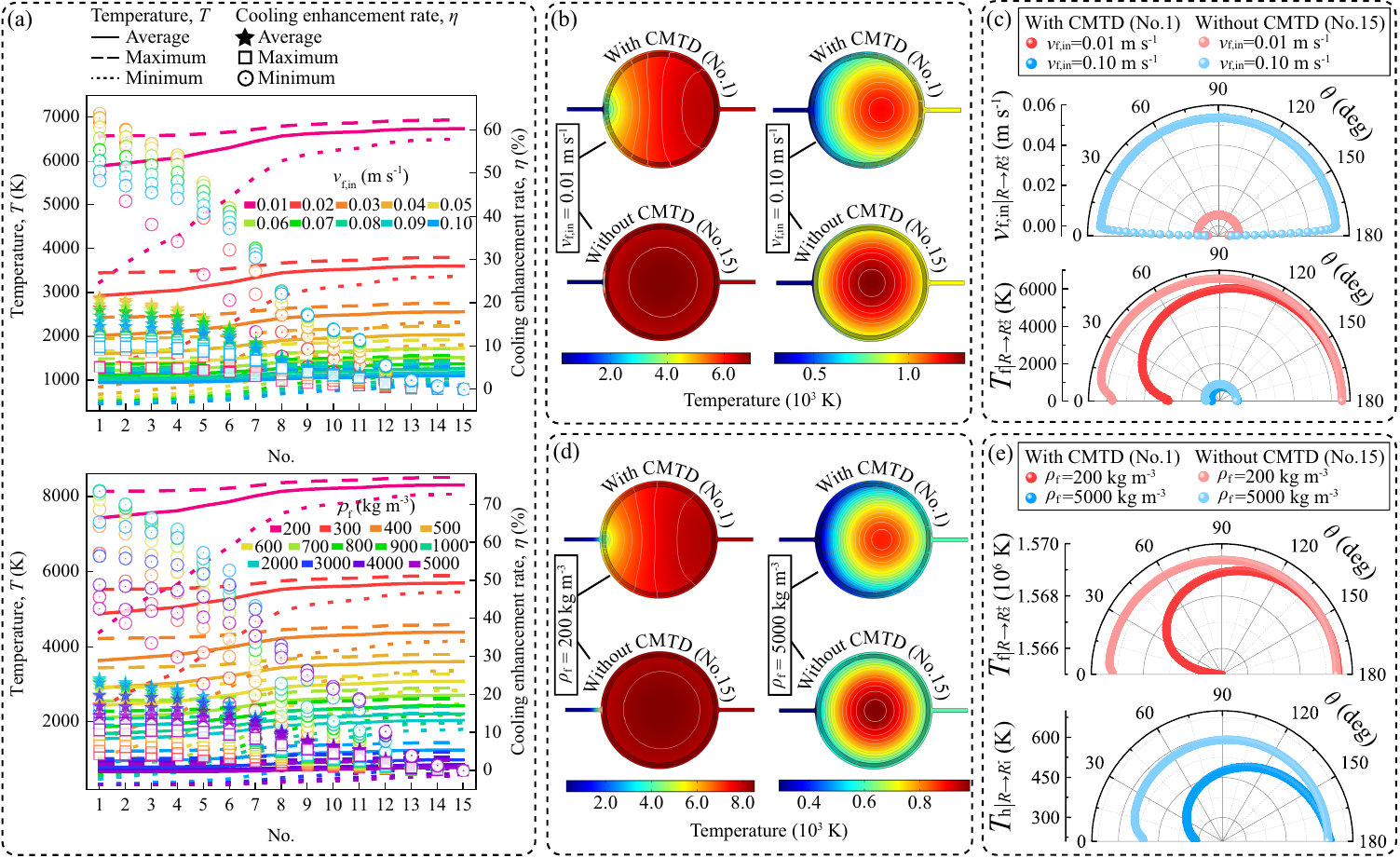}
\caption{\label{Fig. 5} 
\footnotesize{Steady cooling enhancement characteristics under different fluidic parameters. (a) The maximum/average/minimum temperature of the IHS and the maximum/average/minimum cooling enhancement rate of the case with and without CMTD. The upper and lower subgraphs consider the variations of $v_\mathrm{f,in}$ and $\rho_\mathrm{f}$, respectively. (b-e) Temperature distributions of the heat transfer model and the tangential temperature/velocity characterizations. (b,d) Temperature distributions of the heat transfer model with and without CMTD under the conditions $v_\mathrm{f,in}=0.01\  \mathrm{m\  s^{-1}}$, $v_\mathrm{f,in}=0.10\  \mathrm{m\  s^{-1}}$ (b); $\rho_\mathrm{f}=200\  \mathrm{kg\  m^{-3}}$, and $\rho_\mathrm{f}=5000\  \mathrm{kg\  m^{-3}}$ (d). (c) Tangential velocity distributions $(v_\mathrm{f}|_{R\rightarrow R_2^{+}})$ and temperature distributions $(T_\mathrm{f}|_{R\rightarrow R_2^{+}})$ of the working fluid close to the package structure for (b). (e) Tangential temperature distributions $(v_\mathrm{f}|_{R\rightarrow R_2^{+}})$ of the working fluid and the tangential IHS temperature $(T_\mathrm{h}|_{R\rightarrow R_1^{-}})$ close to the package structure for (d).
}}
 \end{figure} 

Due to the conservation of energy, there exists a relationship between the difference in inlet and outlet temperatures of the cooling fluids and the power of the IHS, expressed as: $\Delta T=\pi R_1^2 \dot\phi_\mathrm{h} d_\mathrm{z}/C_\mathrm{f}$. This indicates that the CMTD effects are influenced not only by $C_\mathrm{f}$ but also by the power provided by the IHS. We first examine this by regulating the IHS ratio of the heat transfer model, maintaining $R_2$ while varying $R_1$ from $2.0\ \mathrm{mm}$ to $2.9\ \mathrm{mm}$ (Fig. 4a, lower subgraph). As $R_1$ increases, the IHS ratio improves, leading to higher IHS power and temperature. Two details are emphasized. Firstly, there is a slight decrease in \( \eta_\mathrm{ave} \) when further increasing \( R_1 \) from \( 2.8\ \mathrm{mm} \) to \( 2.9\ \mathrm{mm} \). This can be attributed to the decrease in the region of the package structure, limiting the capacity to regulate heat flow for CMTD. To investigate this, we performed an additional study on the influence of \( \dot\phi_\mathrm{h} \) on cooling enhancement effects under a constant $(R_2-R_1)$, excluding the influence of the package structure region (Fig. 6a). The positive correlation between the cooling enhancement rates and \( \dot\phi_\mathrm{h} \) further confirms that the CMTD effect becomes significant with the increase in the power of the IHS, suggesting the potential of CMTD in high-power-density thermal management areas. Secondly, there is a secular rise in cooling enhancement rates with the increasing IHS power. We examined two extreme conditions to characterize the temperature distributions of the heat transfer model under varying \( R_2 \) and \( \dot\phi_\mathrm{h} \) (Fig. 4d,e and Fig. 6b,c). Under both lower and higher IHS power density, the temperature redistribution effect enabled by CMTD remains similar. The insignificant variation in redistribution effect under different cases leads to the secular rise in cooling enhancement rates with increasing IHS power. Our theoretical analysis further provides a detailed explanation for this result (Supplemental Note 5).

  \begin{figure}[h]
\centering
\includegraphics[width=\textwidth]  {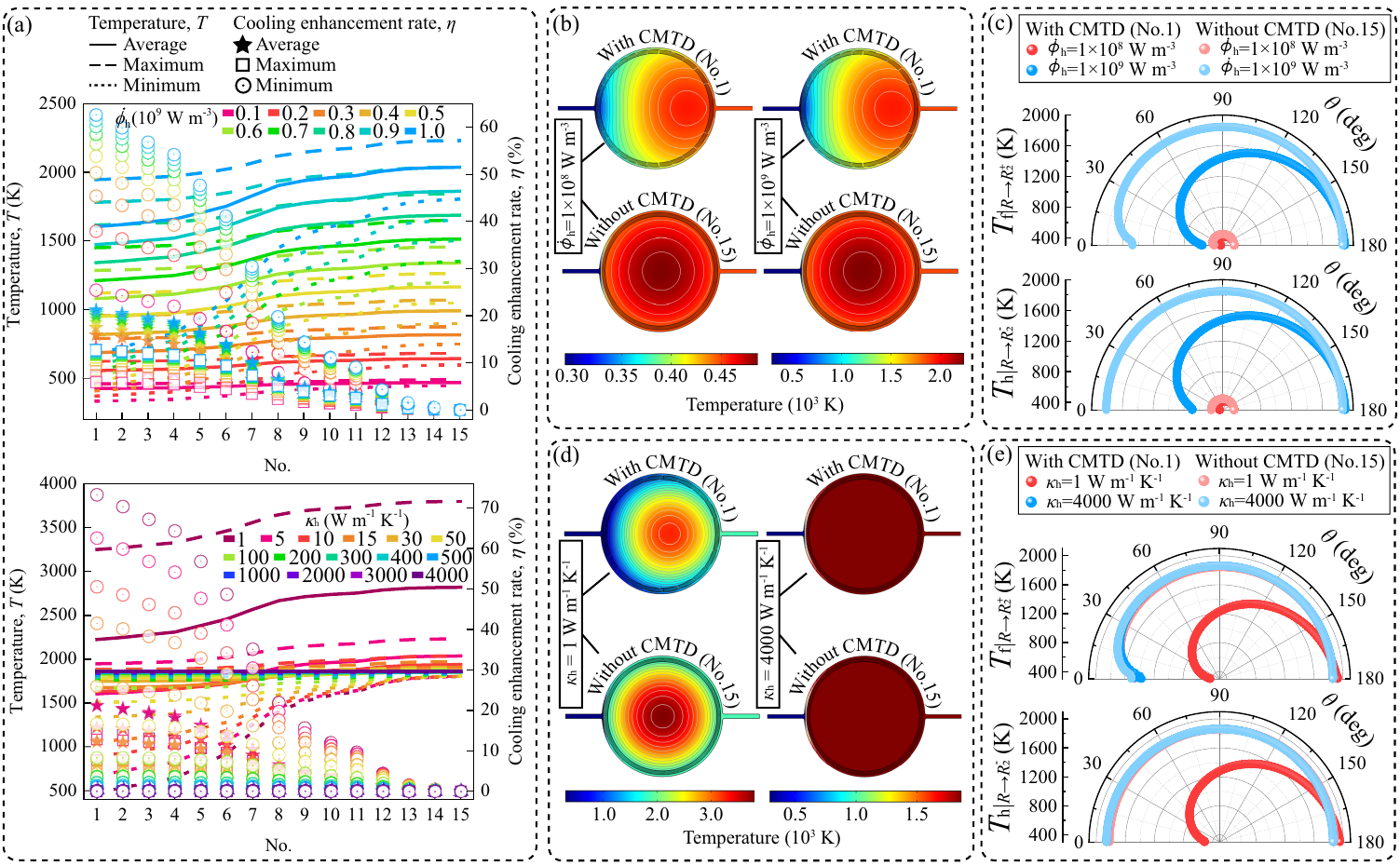}
\caption{\label{Fig. 6}
\footnotesize{Steady cooling enhancement characteristics under different IHS parameters. (a) The maximum/average/minimum temperature of the IHS and the maximum/average/minimum cooling enhancement rate of the case with and without CMTD. The upper and lower subgraphs consider the variations of $\dot\phi_\mathrm{h}$ and $\kappa_\mathrm{h}$, respectively. (b-e) Temperature distributions of the heat transfer model and the tangential temperature/velocity characterizations. (b,d) Temperature distributions of the heat transfer model with and without CMTD under the conditions $\dot\phi_\mathrm{h}=1\times 10^8\  \mathrm{W\  m^{-3}}$, $\dot\phi_\mathrm{h}=1\times 10^9\   \mathrm{W\  m^{-3}}$ (b); $\kappa_\mathrm{h}=1\  \mathrm{W\  m^{-1}\  K^{-1}}$, and $\kappa_\mathrm{h}=4000\  \mathrm{W\  m^{-1}\  K^{-1}}$ (d). (c,e) Tangential temperature distributions of the working fluid $(T_\mathrm{f}|_{R\rightarrow R_2^{+}})$ and IHS $(T_\mathrm{h}|_{R\rightarrow R_2^{-}})$ close to the package structure for (b) [(c)] and (d) [(e)].
}}
 \end{figure} 
 
To capture more mechanisms, we define a factor \( \beta \) to characterize the ease of cooling fluid temperature variation per IHS power, given by \( \beta=C_\mathrm{f}/(\pi R_1^2\dot\phi_\mathrm{h}d_\mathrm{z}) \). We summarize all parameter factors influencing the relationship between \( \eta_\mathrm{ave} \) and \( \beta \) (Fig. 7a). It is found that \( \eta_\mathrm{ave}(\beta) \) induced by $R_1$ and $v_\mathrm{f,in}$ reaches a maximum value of 21.2\% when \( \beta \) is around $0.7\times 10^3\   \mathrm{K}^{-1}$, while it reaches another maximum value of 24.5\% induced by $\rho_\mathrm{c}$ and $c_\mathrm{f}$. These differences are caused by variations in $\kappa_\mathrm{f}$ from $0.5\  \mathrm{W\  m^{-1}\  K^{-1}}$ to $5\  \mathrm{W\  m^{-1}\  K^{-1}}$, indicating that the CMTD effect can be further enhanced by regulating fluidic parameters (Supplementary Table 5). To provide an analytical explanation, we rearrange Eq. (1) and obtain the relationships between \( \eta_\mathrm{ave} \) and thermal/structural parameters
\begin {equation}
\begin{aligned}
\eta_\mathrm{ave}&=\frac{h\kappa_\mathrm{r}R_2\left(\left(36\kappa_\mathrm{h}+\beta\dot\phi_\mathrm{h}\left(-5a^2-18a R_1+R_1^2\right)\right)E\left[\varOmega\right]-\beta\dot\phi_\mathrm{h}\left(a+R_1\right)^2K\left[\varOmega\right]\right)}{6 E\left[\varOmega\right]\left(\kappa_\mathrm{r}\left(6h\kappa_\mathrm{h}R_2+\beta\dot\phi_\mathrm{h}R_1^2\left(3\kappa_\mathrm{h}+hR_2\right)+6\beta h \kappa_\mathrm{h} R_2 T_\mathrm{f,in}\right)+3\beta h\kappa_\mathrm{h}\dot\phi_\mathrm{h}R_1^2R_2\mathrm{ln}\frac{R2}{R1}\right)}\\
&=F\left[\beta \left(c_\mathrm{f}, \rho_\mathrm{f}, v_\mathrm{f,in}, D, R_1, \dot
\phi_\mathrm{h}\right), h\left(c_\mathrm{f}, \rho_\mathrm{f}, v_\mathrm{f,in}, \mu, \kappa_\mathrm{f}, R_2, R_3\right), \dot\phi_\mathrm{h}, \cdots\right]
\end{aligned}
\end{equation}
where $\varOmega=-4aR_1/\left(a-R_1\right)^2$, and \(F\left[\cdots\right]\) denotes a mapping relationship among \( \eta_\mathrm{ave} \), \( \beta \), \( h \), \( \dot\phi_\mathrm{h} \), and several thermal and structural parameters. Since various influencing factors, such as \( v_\mathrm{f,in} \), \( \rho_\mathrm{f} \), and \( c_\mathrm{f} \), simultaneously influence \( \beta \) and \( h \), there exists a suitable combination of these factors for maximizing \( \eta_\mathrm{ave} \). Interestingly, \( \kappa_\mathrm{f} \) only impacts \( h \) and has no influence on \( \beta \). Therefore, the original maximum value of \( \eta_\mathrm{ave} \) is improved by new combinations of \( h \) and other thermal/structural parameters with a slight increase in \( \beta \) (see the region marked by the red ellipse in Fig. 7a, Supplementary Note 6).

Moreover, we notice that $ \dot\phi_\mathrm{h} $ is an independent value that can be separated (Eq. (2)). It implies that enhancing $ \dot\phi_\mathrm{h} $ can further enhance $ \eta_\mathrm{ave} $. When $ \dot\phi_\mathrm{h}\rightarrow \infty $, $ \eta_\mathrm{ave} $ reaches its maximum value. This result agrees with the aforementioned temperature field analysis and is further illustrated in the relationship $ \eta-\beta(\dot\phi_\mathrm{h}) $ (Fig. 7a), indicating the potential of CMTD in thermal management applications with high power density.

Finally, we consider the influence of IHS parameters on the CMTD effect. Since the thermal parameters of the cooling fluid and package structure are determined in the above content, we additionally insert the influence of \( \kappa_\mathrm{h} \) (Fig. 6a, lower subgraph). It is observed that with increasing \( \kappa_\mathrm{h} \), the redistribution effect of cooling fluids enabled by CMTD weakens (Fig. 6d,e). This is because large \( \kappa_\mathrm{h} \) leads to a uniform temperature distribution in the IHS, diminishing the CMTD effect. It suggests that when the thermal conductivity of the IHS is low, CMTD can spontaneously enhance the cooling efficiency. We further generalize the concept of self-adaptive cooling enhancement in the aforementioned relationship \( \eta_\mathrm{ave}(v_\mathrm{f,in}) \) by summarizing various thermal parameters under varying parameters (Fig. 7b). Within a certain range, \( \eta_\mathrm{ave} \) spontaneously enhances with decreasing thermal parameters. This generalized spontaneous cooling enhancement breaks the consensus in existing thermal management techniques and demonstrates that the significant effect of CMTD plays a significant role when the thermal parameters are weak, paving the way for its wide applications.

  \begin{figure}[h!]
\centering
\includegraphics[width=\textwidth]  {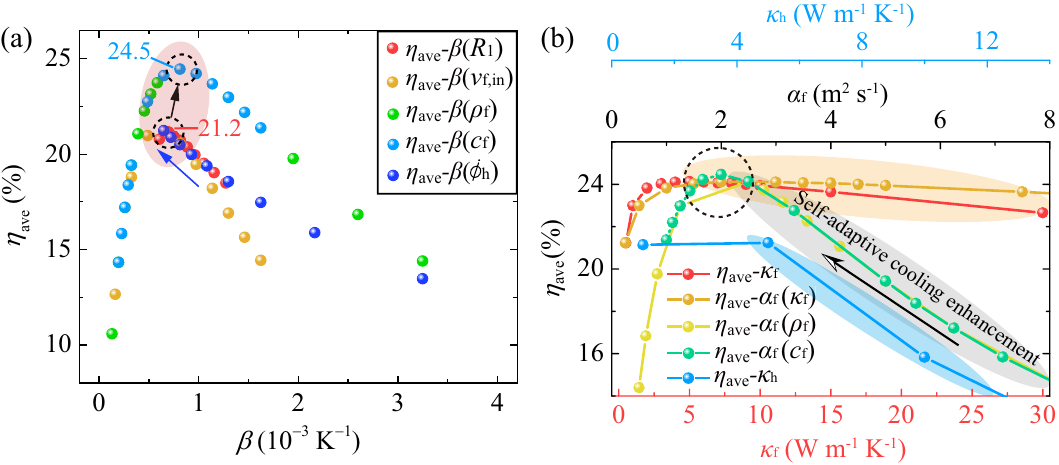}
\caption{\label{Fig. 7}
\footnotesize{
Optimization mechanism and generalized interpretation of self-adaptive cooling enhancement. (a) Relationships between the average cooling enhancement rate $(\eta_\mathrm{ave})$ and the ratio of heat capacity flow rate to IHS power $(\beta)$ under different variations of parameters. Note that variations in $\beta$ are induced by five factors: $R_1$, $v_\mathrm{f,in}$, $\rho_\mathrm{f}$, $c_\mathrm{f}$, and $\dot\phi_\mathrm{h}$. (b) Relationships between $\eta_\mathrm{ave}$ and various thermal parameters, including $\kappa_\mathrm{f}$, $\kappa_\mathrm{h}$, and $\alpha_\mathrm{f}$, where $\alpha_\mathrm{f}$ denotes the thermal diffusivity of the cooling fluid. Note that variations in $\alpha_\mathrm{f}$ are induced by three factors: $\kappa_\mathrm{f}$, $\rho_\mathrm{f}$, and $c_\mathrm{f}$.
}}
 \end{figure}

\subsection{Transient cooling enhancement characteristics with non-trivial heat transfer enhancement}

To investigate the CMTD effect in a transient heat transfer scenario, we kept the structural and fluidic parameters consistent with the steady-state heat transfer model shown in Fig. 5a ($v_\mathrm{f,in}=0.04\ \mathrm{m\ s^{-1}}$). The initial temperature of the IHS ($T_\mathrm{h,in}$) was varied from 3000 K to 5000 K in intervals of 500 K, while maintaining the initial temperature of other components at 293.15 K. Subsequently, simulations were conducted to produce temperature-time curves for the IHS, demonstrating its cooling performance, as depicted in Fig. 8a. For clarity, temperature distributions of the heat transfer model at characteristic time points are illustrated in Fig. 8b.

\begin{figure}[h!]
\centering
\includegraphics[width=\textwidth]  {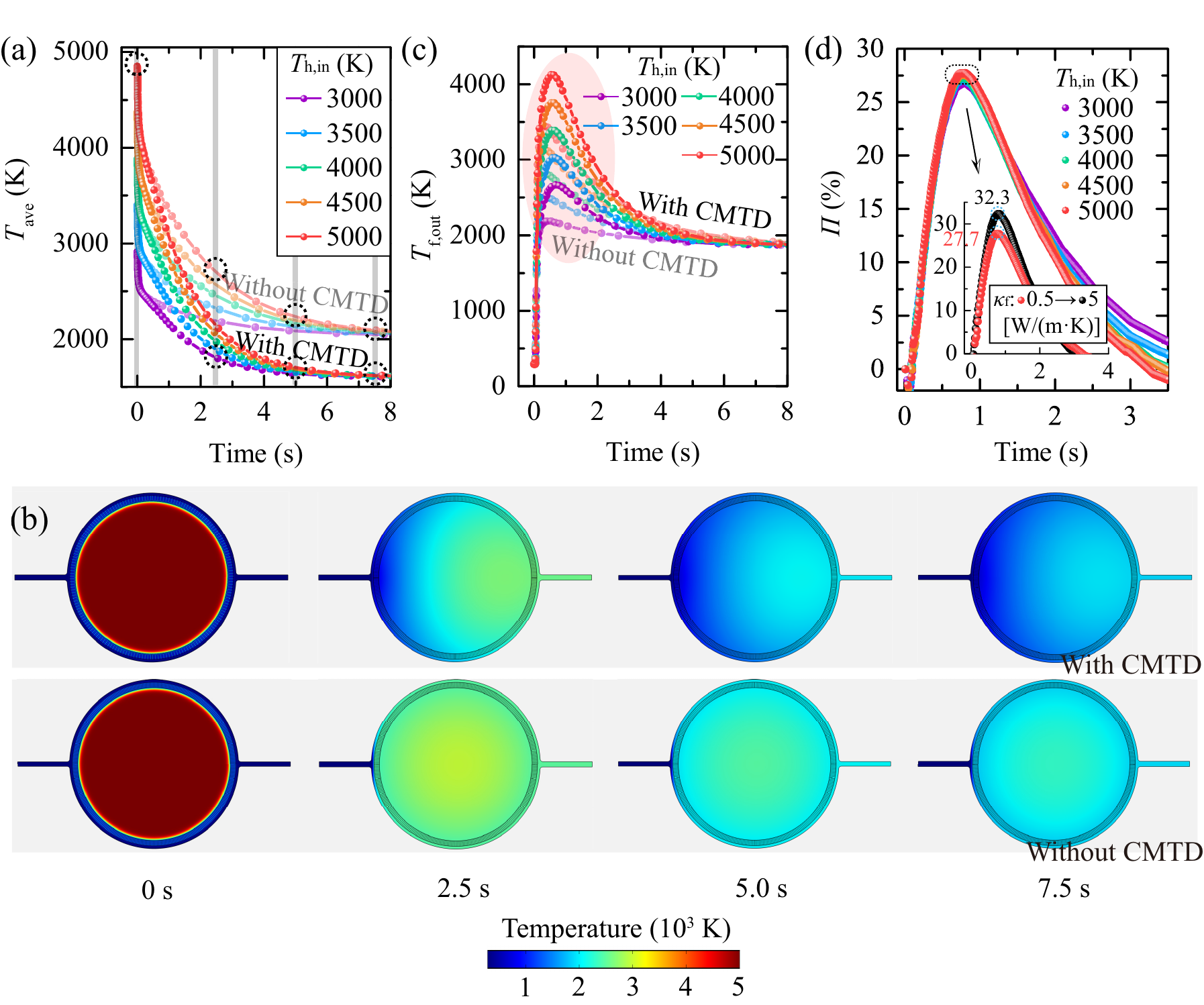}
\caption{\label{Fig. 8}
\footnotesize{Transient cooling of the IHS with different initial temperatures ($T_\mathrm{h,in}$). (a) The average temperature of the IHS $(T_\mathrm{ave})$ varied with time under different $T_\mathrm{h,in}$. (b) Temperature distributions of the heat transfer model with and without CMTD under $T_\mathrm{h,in}=5000\ \mathrm{K}$ at characteristic times: 0 s, 2.5 s, 5.0 s, and 7.5 s. Note that the $T_\mathrm{ave}$ corresponding to each subgraph is marked in (a). (c) Outlet temperature of the working fluid $(T_\mathrm{f,out})$ varied with time under different $T_\mathrm{h,in}$. (d) Heat transfer enhancement degree $(\varPi)$ enabled by CMTD varied with time under different $T_\mathrm{h,in}$. Refer to Supplementary Fig. 7 for more details on the transient cooling characteristics under $\kappa_\mathrm{f}=5.0\ \mathrm{W\ m^{-1}\ K^{-1}}$.
}}
 \end{figure}

The utilization of CMTD accelerates the cooling process of the IHS by effectively enhancing heat exchange between the IHS and the cooling fluid during transient heat transfer. This enhancement can be indirectly assessed by analyzing the outlet temperature of the cooling fluid, as illustrated in Fig. 8c. Quantitative assessment of the enhanced heat transfer effect achieved by CMTD is detailed in Fig. 8d (Methods), demonstrating a maximum heat transfer enhancement degree $(\varPi)$ increase of up to 27.7\%. As mentioned earlier, regulating $h$ can enhance the CMTD effect. Introducing the impact of $\kappa_\mathrm{f}$ by varying it from $0.5\ \mathrm{W\ m^{-1}\ K^{-1}}$ to $5\ \mathrm{W\ m^{-1}\ K^{-1}}$, the maximum $\varPi$ under $\kappa_\mathrm{f}=5\ \mathrm{W\ m^{-1}\ K^{-1}}$ is found to be $32.3\%$, showcasing the effectiveness of CMTD under transient cooling scenarios.

Further examination of the influence of the initial temperature of the IHS on the enhanced heat transfer effect reveals minimal impact on the degree of maximum heat transfer enhancement. Interestingly, as the initial temperature decreases, the heat transfer enhancement effect of CMTD persists for a longer duration. This observation suggests that under limited heat transfer temperature differences – conditions typically considered unfavorable for heat transfer enhancement – CMTD exhibits significant application potential. These findings are consistent with the self-adaptive characteristics observed under steady-state heat transfer, highlighting the efficacy of CMTD under restrictive heat transfer conditions with low heat transfer temperature and thermal parameters.

Remarkably, the heat transfer enhancement achieved by CMTD during transient cooling processes does not necessitate improvements in thermal parameters or surface enlargement, as common technologies require. The mechanism illustrated in Fig. 2a also operates in transient heat transfer situations, redistributing the tangential fluid temperature and achieving higher heat transfer rates in the initial cooling states (Fig. 8b, 2.5 s). This nontrivial heat transfer enhancement mechanism elucidates the heat transfer relationship in CMTD strategy in-depth and guides practical, efficient thermal management in real-world applications.

\section{Conclusions}

In this study, we introduced the concept of convective meta-thermal dispersion (CMTD) for self-adaptive cooling enhancement, demonstrating that hybridizing low thermal conductance into high thermal conductivity package structures optimizes convective heat transfer between the internal heat source (IHS) and the cooling fluid. By adjusting the heat flow inside the package region, employing equivalent radial high thermal conductivity and equivalent tangential low thermal conductivity, we altered the temperature distribution of the fluid inside the flow channel, ultimately enhancing IHS cooling, with a maximum reduction in the average temperature of the IHS by 24.5\%.

We identified a positive correlation between the cooling enhancement effect facilitated by CMTD and the power density of the IHS, indicating significant promise for effective thermal management in high-power-density applications such as nuclear fission and fusion. At ultra-high power levels, where the heat generated by reactions is absorbed by the fluid for utilization, efficiently adjusting the fluid temperature across orders of magnitude effectively regulates the IHS temperature of the reaction area (Fig. 6c).

Furthermore, transient simulations were utilized to study the enhancement of heat exchange ability between the IHS and working fluid by CMTD. Results indicated a significantly higher cooling rate for an IHS using CMTD compared to one without, with a maximum heat transfer enhancement up to 32.3\%. Combining steady and transient results, CMTD demonstrated significant effects on spontaneously enhancing IHS cooling performance under limited heat transfer conditions, including decreasing inlet velocity, components' thermal conductivities/diffusivities, and heat transfer temperature differences, displaying self-adaptive cooling enhancement characteristics.

Interestingly, recent research has highlighted that defect scattering can enhance thermal conductance by restoring the directional equilibrium of phonons in nanoscale heating zones \cite{hu2024defect}. Our findings suggest that optimizing convective heat transfer can be achieved by introducing low thermal conductance to regulate heat flow from the perspective of macroscopic heat transfer. These unusual insights might deepen the understanding of heat transfer mechanisms.

Compared to existing advanced heat transfer enhancement technologies, the CMTD approach does not require enhancing thermal properties or enlarging heat transfer areas, thus conserving energy and materials and promoting sustainability. For practical applications, package structures with CMTD can be fabricated by drilling holes in the packaging region with high thermal conductivity while ensuring mechanical properties, similar to obtaining shells with anisotropic thermal conductivity in thermal metamaterials fabrication \cite{huang2020theoretical, xu2023transformation, YangS, zhang2023diffusion, yang2023controlling}. Our findings provide design concepts and a working mechanism for efficient thermal management systems while exploring heat transfer enhancement mechanisms in depth.

In future research, cooling enhancement could be extended to heating operations to achieve heating enhancement, replacing cooling conditions and internal heat sources with heating conditions and internal cold sources, respectively. This presents opportunities for various applications, such as improving Stirling engine efficiency by increasing temperature differences between its hot and cold ends \cite{zhou2023convective}. Additionally, advancements in energy storage technologies could incorporate thermal/cold energy storage materials as internal heat and cold sources, potentially enhancing energy storage/release efficiency \cite{woods2021rate, aftab2021phase}. Furthermore, the diffusion-based concept of heat transfer could be expanded to mass transfer \cite{zhang2023diffusion, yang2023controlling}, potentially enhancing system efficiency in various diffusion systems, such as humidity control \cite{li2019full}, moisture sorption-desorption \cite{wang2020thermal}, gas separation \cite{chen2021tailoring}, and plasma technology \cite{zhang2022transformation}, offering potential benefits for energy and mass control technologies across diverse energy sectors.

\section*{Methods}

\subsection*{Finite element simulations}

To investigate the cooling enhancement effect of CMTD, the cooling fluid is assumed to be incompressible, viscous, and with constant properties. The convective heat transfer in the cooling fluid, under a steady state, is governed by the following equations:

\begin {equation}
\left\{\begin{aligned}
&\nabla \cdot \boldsymbol v =0\\
&\rho_\mathrm{f}(\boldsymbol v \cdot \nabla) \boldsymbol v=\rho_\mathrm{f} \boldsymbol F-\nabla p+\mu \nabla^2\boldsymbol v \\
&\boldsymbol v \cdot \nabla T_\mathrm{f}=\frac{\kappa_\mathrm{f}}{\rho_\mathrm{f}c_\mathrm{f}}\nabla^2T_\mathrm{f}+\dot \varPhi_\mathrm{f}
\end {aligned}
\right.,
\end {equation}
Here, $\boldsymbol v$ is velocity, $\rho_\mathrm{f}$ is density, $\boldsymbol F$ is body force, $p$ is pressure, $\mu$ is dynamic viscosity, $\kappa_\mathrm{f}$ is thermal conductivity, $c_\mathrm{f}$ is heat capacity, $T_\mathrm{f}$ is temperature, and $\dot \varPhi_\mathrm{f}$ is the heat source power due to viscous dissipation, all representing parameters of the cooling fluid.

The conductive heat transfer in the package and IHS regions, both with constant properties, under a steady state, is described by the following equations:

\begin{equation}
\left\{
\begin{aligned}
&\nabla^2 T_\mathrm{s} = 0\\
&\nabla^2 T_\mathrm{h} + \frac{\dot\phi_\mathrm{h}}{\kappa_\mathrm{h}} = 0
\end{aligned}
\right.,
\end{equation}
where $T_\mathrm{s}$ is the temperature distribution in the package region, $T_\mathrm{h}$ is the temperature distribution in the IHS region, and $\kappa_\mathrm{h}$ is the thermal conductivity of IHS. 

The heat transfer model is numerically solved with boundary conditions as depicted in Fig. 1e. The inlet temperature of the working fluid $T_\mathrm{f,in}$ was 293.15 K, and the outlet pressure $p_\mathrm{out}$ was 101325 Pa (absolute pressure). The upper and lower walls were insulated. The model was then solved to study the influence of various parameters on the CMTD.

For the model in Fig. 1b,c, $\kappa_\mathrm{s}=20\  \mathrm{W\  m^{-1}\  K^{-1}}$ is referenced from the thermal conductivity of stainless steel, while $\kappa_\mathrm{s}=4000\  \mathrm{W\  m^{-1}\  K^{-1}}$ is referenced from the thermal conductivity of graphene. The distinction between strong and weak convection is defined based on the temperature variation of the cooling fluid along the flow channel. Strong convection is characterized by an outlet temperature approach to the inlet temperature, while weak convection is characterized by a significantly higher outlet temperature than the inlet temperature. The fluid parameters were as follows: thermal conductivity $\kappa_\mathrm{f}=0.5\  \mathrm{W\  m^{-1}\  K^{-1}}$, density $\rho_\mathrm{f}=1000\  \mathrm{kg\  m^{-3}}$, specific heat capacity $c_\mathrm{f}=2000\  \mathrm{J\  kg^{-1}\  K^{-1}}$, specific heat ratio $\gamma=1$, and dynamic viscosity $\mu=0.01\   \mathrm{Pa\  s}$. The shell parameters were: thermal conductivity $\kappa_\mathrm{s}=20\  \mathrm{W\  m^{-1}\  K^{-1}}$ or  $\kappa_\mathrm{s}=4000\  \mathrm{W\  m^{-1}\  K^{-1}}$. The IHS parameters were: thermal conductivity $\kappa_\mathrm{h}=5\  \mathrm{W\  m^{-1}\  K^{-1}}$ and power density $\dot\phi_\mathrm{h}=1\times10^9\  \mathrm{W\  m^{-3}}$. For strong convective heat transfer in Fig. 1b, the solid-fluid heat transfer and turbulent flow modules ($k-\varepsilon$) were employed. The boundary conditions include an inlet velocity $v_\mathrm{f,in}=40\  \mathrm{m\  s^{-1}}$, inlet temperature $T_\mathrm{f,in}=273.15\  \mathrm{K}$, and outlet pressure (absolute pressure) $p_\mathrm{out}=101325\  \mathrm{Pa}$. For weak convective heat transfer in Fig. 1c, the fluid-solid heat transfer and laminar flow modules were used. In this case, $v_\mathrm{f,in}$ was reset to $0.004\  \mathrm{m\  s^{-1}}$, while other conditions remained the same as those of strong convective heat transfer. For the model in Figs. 4-6, the simulation methods are similar to that of Fig. 1. The operating conditions of the model in Fig. 2 are identical to those in Fig. 5a. The different operating conditions are listed in Supplementary Tables 4-6. For the transient simulations of Fig. 8, the density and specific heat capacity of the IHS and package structure were set as $1000\  \mathrm{kg\  m^{-3}}$ and $1000\  \mathrm{J\  kg^{-1}\  K^{-1}}$, respectively. The other parameters were the same as those in the upper subgraph of Fig. 5a ($v_\mathrm{f,in}=0.04\  \mathrm{m\   s^{-1}}$).

\subsection*{Definition of anisotropic degree within the package structure}

The equivalent radial and tangential thermal conductivities of the package structure are given by
\begin{equation}
\kappa_\mathrm{r}=\frac{\kappa_\mathrm{A}+\kappa_\mathrm{B}}{2},
\end{equation}
and
\begin{equation}
\kappa_\mathrm{t}=\frac{2\kappa_\mathrm{A}\kappa_\mathrm{B}}{\kappa_\mathrm{A}+\kappa_\mathrm{B}},
\end{equation}
respectively (Supplementary Note 4). Then, the anisotropic degree of the thermal conductivity of the package structure is defined as
\begin{equation}
\varXi=\frac{\kappa_\mathrm{r}}{\kappa_\mathrm{t}}.
\end{equation}

\subsection*{Definition of cooling enhancement rates}
The finite element simulation encompasses both steady and transient states. For steady-state simulation, its expression represents the temperature distribution of the heat transfer model. To quantitatively study the effect of CMTD on the temperature reduction of the IHS, three cooling efficiencies were defined based on the minimum, average, and maximum temperature of the IHS.

Firstly, the average cooling efficiency is given by:

\begin{equation}
\eta_\mathrm{ave} = \frac{T_{\mathrm{ave},i}|_{i=15}-T_{\mathrm{ave},i}|_{i=1\cdot\cdot\cdot15}}{T_{\mathrm{ave},i}|_{i=15}} \times 100\%,
\end{equation}
where $i$ in $T_{\mathrm{ave},i}$ denotes the number of the combinations of $\kappa_\mathrm{A}$ and $\kappa_\mathrm{B}$ of the package structure. As mentioned earlier, $i=15$ ($\kappa_\mathrm{A} = \kappa_\mathrm{B}$) corresponds to a classical high thermal conductivity structure. Therefore, the definition in Eq. (8) essentially characterizes the reduction effect of different levels of equivalent anisotropy on the average temperature decrease of IHS relative to the classical high thermal conductivity package structure.

Similarly, we define:

\begin{equation}
\eta_\mathrm{max} = \frac{T_{\mathrm{max},i}|_{i=15}-T_{\mathrm{max},i}|_{i=1\cdot\cdot\cdot15}}{T_{\mathrm{max},i}|_{i=15}} \times 100\%
\end{equation}
and

\begin{equation}
\eta_\mathrm{min} = \frac{T_{\mathrm{min},i}|_{i=15}-T_{\mathrm{min},i}|_{i=1\cdot\cdot\cdot15}}{T_{\mathrm{min},i}|_{i=15}} \times 100\%.
\end{equation}
The former and the latter assess the cooling enhancement performance of the CMTD on the maximum and minimum temperatures of the IHS, respectively.

\subsection*{Calculation of heat transfer rate and heat transfer enhancement degree}
For transient simulations, the primary objective is to characterize the impact of CMTD on the cooling effect of IHS. Specifically, this involves assessing changes in the amount of heat flow per unit of time transferred from the IHS to the working fluid. According to the principle of energy conservation, the amount of heat absorbed by the working fluid per unit of time is equal to the amount of heat transferred from the IHS to the working fluid.

Initially, by setting the inlet temperature of the working fluid, $T_\mathrm{f,in}$, to a constant value, the real-time outlet temperature $T_\mathrm{f,out}(t)$ can be obtained through boundary probes in COMSOL Multiphysics. This allows the calculation of the heat absorbed/heat transferred rate:

\begin{equation}
Q_u(t) = c_\mathrm{f}v_\mathrm{in}Dd_\mathrm{z}\rho_\mathrm{f}\left[T_\mathrm{f,out}(t) - T_\mathrm{f,in}\right],
\end{equation}
where $u$ represents the structure of the packaging region, and $u=\mathrm{2}$ and $u=\mathrm{1}$ represent the case with and without CMTD, respectively. This approach enables the definition of the degree of heat transfer enhancement at any given time:

\begin{equation}
\varPi(t) = \frac{Q_2(t) - Q_1(t)}{Q_1(t)} \times 100\%.
\end{equation}
\subsection*{Grid independence test}
We conducted corresponding grid independence analyses to ensure the accuracy of the simulations. Firstly, we set up four cases for the model in Fig. 1e (Supplementary Table 7). The remaining parameters were set as follows: the working fluid parameters $\kappa_\mathrm{f}=0.5\ \mathrm{W\ m^{-1}\ K^{-1}}$, $\rho_\mathrm{f}=1000\ \mathrm{kg\ m^{-3}}$, $c_\mathrm{f}=2000\ \mathrm{J\ kg^{-1}\ K^{-1}}$, $\mu=0.01\ \mathrm{Pa\ s}$, $p_\mathrm{out}=101325\ \mathrm{Pa}$, and $T_\mathrm{f,in}=293.15\ \mathrm{K}$; the structural parameters $R_1=2.8\ \mathrm{mm}$, $R_2=3.0\ \mathrm{mm}$, $D=0.2\ \mathrm{mm}$, $L=5\  \mathrm{mm}$ and $\omega=1^{\circ}$; the IHS parameters $\kappa_\mathrm{h}=5\ \mathrm{W\ m^{-1}\ K^{-1}}$. Based on these parameters, finite element simulations can be performed for the four specified cases. Here, the fluid outlet temperature was selected as the reference data. According to the conservation of energy, the outlet temperature of the fluid under steady-state conditions was obtained as:

\begin{equation}
T_\mathrm{f,out}=T_\mathrm{f,in}+\frac{\pi R_1^2 \dot\phi}{c_\mathrm{f}\rho_\mathrm{f}v_\mathrm{f,in}D}.
\end{equation}
We compared the calculated theoretical values with the simulated values and calculated their deviations:

\begin{equation}
E=\frac{T_\mathrm{out}^\mathrm{sim}-T_\mathrm{out}^\mathrm{ana}}{T_\mathrm{out}^\mathrm{ana}}\times 100\%,
\end{equation}
where $T_\mathrm{out}^\mathrm{sim}$ and $T_\mathrm{out}^\mathrm{ana}$ represent the outlet temperature of the working fluid calculated through simulation and analytical theory, respectively. The calculation results are shown in Supplementary Table 8. It can be observed that as the number of grids increases, the relative deviation gradually decreases. Further, six parameters—$T_\mathrm{ave}$, $T_\mathrm{max}$, $T_\mathrm{min}$, $\eta_\mathrm{ave}$, $\eta_\mathrm{max}$, and $\eta_\mathrm{min}$—were selected. The changes in these six parameters under different grid schemes are shown in Supplementary Fig. 8. It can be observed that, regardless of how the grid changes, it does not affect the trend of these parameters. From the perspective of accuracy, the results obtained from grid schemes 3-5 are relatively close. Considering computational efficiency and accuracy, all simulations in this work adopt scheme 3.

\section*{Authorship contribution}
\textbf{Xinchen Zhou:} Conceptualization, Data curation, Formal analysis, Investigation, Methodology, Software, Validation, Visualization, Writing - original draft, review \& editing. \textbf{Ruzhu Wang:} Conceptualization, Resources, Writing - review. \textbf{Xiaoping Ouyang:} Conceptualization. \textbf{Jiping Huang:} Conceptualization, Supervision, Funding acquisition, Resources, Project administration, Writing - review \& editing.

\section*{Acknowledgements}
We thank Ms. Fangfang Deng, and Mr. Zhao Shao for their beneficial discussions. We gratefully acknowledge funding from the National Natural Science Foundation of China (Grants No. 12035004 and No. 12320101004) and the Innovation Program of Shanghai Municipal Education Commission (Grant No. 2023ZKZD06).
\bibliographystyle{unsrt}
\bibliography{mybib}

\end{document}